\pdfoutput=1
\documentclass[reprint,pre,amsmath,superscriptaddress,amssymb,
 aps]{revtex4-1}
\usepackage{graphicx,subfigure,color}
\begin{document}

\title{Disorder and the neural representation of complex odors: \\ smelling in the real world}
\date{\today}

\author{Kamesh Krishnamurthy}
\thanks{co-first author}
\affiliation{David Rittenhouse and Richards Laboratories, University of Pennsylvania, Philadelphia, PA 19104, USA}

\author{Ann M. Hermundstad}
\thanks{co-first author}
\affiliation{Janelia Research Campus, Howard Hughes Medical Institute, 19700 Helix Drive, Ashburn, VA 20147, USA}

\author{Thierry Mora}
\affiliation{Laboratoire de physique statistique, UMR8550, CNRS, UPMC and \'Ecole Normale Sup\'erieure, Paris 75005, Paris}

\author{Aleksandra M. Walczak}
\affiliation{Laboratoire de Physique Th\'eorique, UMR8549m CNRS, UPMC and \'Ecole Normale Sup\'erieure, Paris 75005, Paris}

\author{Vijay Balasubramanian}
\thanks{lead author}
\affiliation{David Rittenhouse and Richards Laboratories, University of Pennsylvania, Philadelphia, PA 19104, USA}

\pacs{}
\keywords{}

\begin{abstract}
Animals 
smelling in the real world
use a small number of receptors to sense a vast number of natural molecular mixtures, and proceed to learn arbitrary associations between odors and valences.  Here, we propose a new interpretation of how the architecture of olfactory circuits is adapted
 to meet these immense complementary challenges.
  First, the diffuse binding of receptors to many molecules compresses a vast odor space into a tiny receptor space, while preserving similarity.  Next, lateral interactions ``densify'' and decorrelate the response, enhancing robustness to noise.  Finally, disordered projections from the periphery to the central brain reconfigure the densely packed information into a format suitable for flexible learning of associations and valences.  We test our theory empirically using data from Drosophila.   Our theory suggests that the neural processing of olfactory information differs from the other senses in its fundamental use of disorder.
\end{abstract}

\maketitle

Animals sense and respond to volatile molecules that carry messages from and about the world.   Some kinds of olfactory behaviors require  sensing of particular molecules such as pheromones.   These molecules and the receptors that bind to them have likely co-evolved over long periods of time to ensure precise and specific binding.  However, to be useful as a general purpose tool  for interaction with a diverse and changing world, the olfactory system should  be prepared to sense and process any volatile molecule.  There are a very large number of such monomolecular odorants (perhaps billions \cite{yu2015drawing}), far more than the number of receptor types available to bind these odorants.  Humans and mice, for instance, have just $\sim300$ and $\sim1000$  functional olfactory receptor types, respectively.  Yet,  animals may be able to discriminate between orders of magnitude more odors than the number of receptor types (a high estimate is given in \cite{Bushdid2014}, but see \cite{10.7554/eLife.08127}).

 At an abstract level, the early stage of the olfactory system faces the immense challenge of embedding a very high-dimensional input space (the space of odor molecules) into a low-dimensional space of sensors (the response space of olfactory receptors).  This embedding must preserve similarity between different odors well enough to permit the judgements of sameness and difference that are crucial for behavior.   Furthermore, experiments   \cite{Choi2011}  suggest that this odor representation is reorganized in higher brain regions to be enormously flexible, allowing learning of nearly arbitrary associations between valences and different groups of odors.    Here, we propose a new theoretical framework (Fig.~1), and provide empirical evidence, suggesting that the olfactory system powerfully exploits physiological and  structural {\it disorder} at different stages of processing to meet these two complementary challenges: ({\it i}) compression of a vast odor space into a tiny receptor space, and ({\it ii}) reorganization of the information to allow flexible learning.

To perform effectively within its design constraints, a sensory system must exploit structure in the environment.  For example, the statistics of natural images dictate an efficient decomposition into edges \cite{olshausen1996emergence},  likely explaining why simple and complex cells in the visual cortex respond preferentially to oriented lines \cite{hubel1962receptive}.    We noted \cite{krishnamurthyCOSYNE14} that a salient feature of natural odors is that they typically contain only a tiny fraction of the possible volatile molecular species.    For example, food odors typically are composed of 3-40 molecules \cite{yu2015drawing}.    Natural odors are thus \emph{sparse} in the high-dimensional space of odorant molecules.  Surprising results from the mathematical literature on random projections \cite{baraniuk2010low,donoho2006compressed,candes2006robust} show that there is an efficient solution for storing signals of this nature: sparse, high-dimensional input signals can be encoded by a compact set of sensors through diffuse and disordered measurements of the input space.  For example, this sort of compression can be achieved if each sensor response contains randomly weighted contributions from every dimension of the input space.    Importantly, this diffuse sensing need not be tuned to the specific structure of the input signal -- i.e. in this manner, it can be non-adaptive.  We propose that the olfactory system employs such a diffuse sensing strategy in order to exploit the sparse structure of natural odor space and produce compact representations of  odors (Fig.~1).

Ultimately, these odor representations must support associations between odors and valence, and experimental evidence suggests that animals can learn such associations both flexibly and reversibly \cite{Choi2011}. However, as we will show, the compact representations achieved by diffuse sensing make such learning difficult.  We show that another form of disorder---a ``densification'' and decorrelation of responses, followed by a disordered expansion---can reorganize odor information into a format that facilitates flexible learning.

We provide evidence for our proposal by analyzing the olfactory system of {\it Drosophila}. We show that the diffuse responses of olfactory receptor neurons provide a compact representation of odor information. We then show that the nonlinear transformation in the second stage of olfactory processing (Antennal Lobe in insects; Olfactory Bulb in mammals), followed by the apparently disordered, expansive projection to the third stage of olfactory processing (Mushroom Body in insects; Piriform Cortex in mammals), facilitate flexible learning of odor categories from small and arbitrarily-chosen groups of sparsely firing neurons. Finally, we demonstrate that the disorder introduced by both the densification and the expansion is critical for robustness to noise.

\begin{figure*}
\begin{center}
\includegraphics[width=0.95\textwidth]{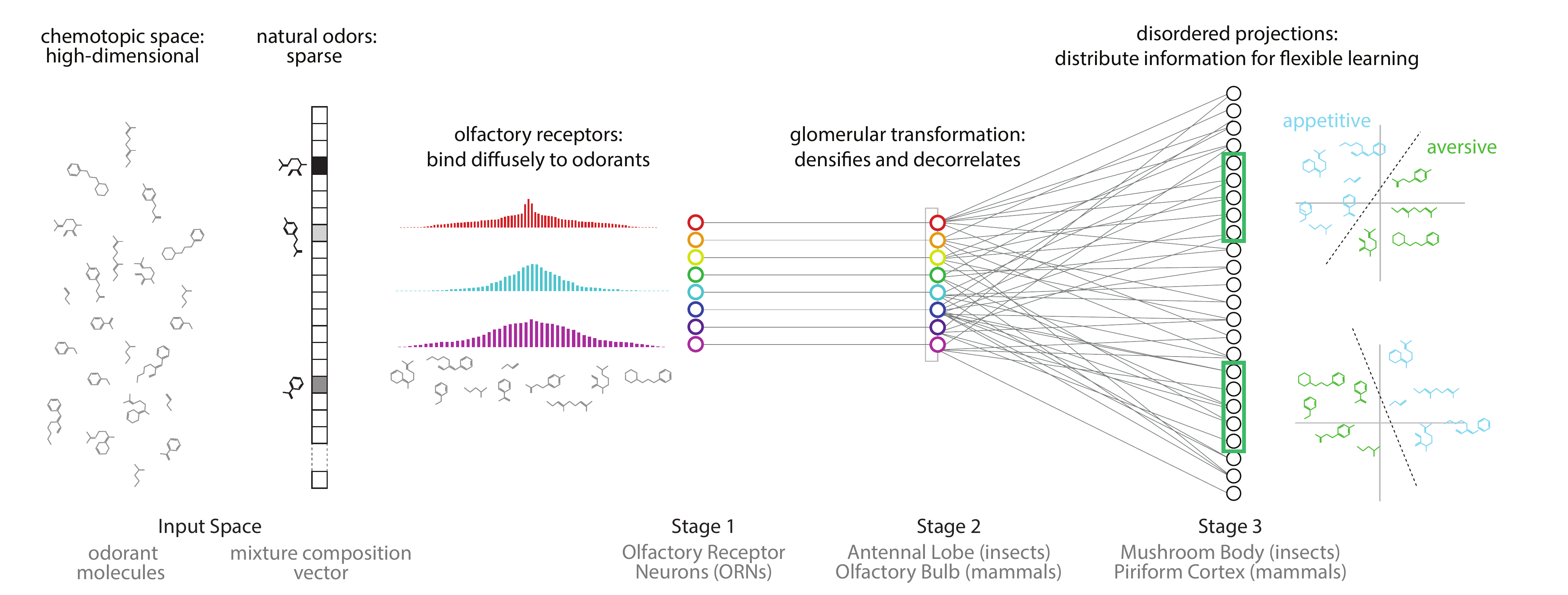}
\end{center}
\caption{
 {\bf Proposal: The olfactory system uses two kinds of disorder to first compress odor information into a small number of receptors, and then reconfigure this information to enable flexible associations between odors and valences.}
 (i) Natural odors are high dimensional but sparse: they contain a tiny fraction of all possible monomolecular odorants.
   (ii) Olfactory receptors diffusely bind
    to a broad range of odorants, producing a compact representation of odor information that enables accurate decoding.
    (iii)   The Antennal Lobe/Olfactory Bulb ``densifies'' and decorrelates this representation, providing robustness to noise.
       (iv)  Disordered projections from the Antennal Lobe/Olfactory Bulb to the Mushroom Body/Piriform Cortex, followed by nonlinearities, create a sparse and distributed representation of odors that facilitates flexible learning of odor categories from small and arbitrarily-chosen subsets of neurons. }\label{fig:Figure_1}
\end{figure*}

\section{Results}

\subsection{Olfactory receptor neurons use disorder to encode natural odors}
Volatile molecules are sensed when they bind to olfactory receptors, each encoded by a separate gene \cite{buck1991novel}. For example, in mice, almost $5\%$ of the genome is devoted to encoding about $1000$ receptor types. Despite such large genomic investments, the number of receptor types is dwarfed by the number of volatile molecules that a general purpose olfactory system might seek to sense.    This raises two related questions.  First, is it possible, even in principle, to sense the high-dimensional space of molecules using the inevitably low-dimensional space of receptor responses?  Second, can this sensing be done by neurons so that odors with similar  mixture compositions  are mapped to nearby regions in response space?

To solve this problem, there is a key simplification that the nervous system could exploit -- natural odors typically contain a tiny fraction of the possible volatile molecules  \cite{yu2015drawing}.  Thus, the representation of a natural odor in terms of its molecular concentration vector is extremely sparse.   Suppose there are $N$ types of volatile molecules, and any given natural odor contains no
more than $K \ll N$ of these types. 
Then, recent results in mathematics show that a small number of linear sensors  (about  $K$) could store complete information about natural odors, provided that their binding affinities were statistically random \cite{baraniuk2010low,donoho2006compressed,candes2006robust}.   This fact suggests a new perspective on the olfactory system:  rather than having strong responses for a specific set of important molecules, a general purpose receptor repertoire should be selected to have molecular affinities that are as disordered as possible, subject to constraints imposed by biophysics and evolution.  Likewise, the quality of olfaction as a general purpose sense will be determined by the degree of disorder in response patterns.

Is there evidence for this view?   Indeed, most Olfactory Receptor Neuron (ORN) types respond diffusely to many odorants, and most odorants evoke diffuse responses from diverse ORN types (insect:  \cite{Hallem2006,carey2010odorant};  mammal: \cite{saito2009odor}).   To assess the quality of the representation  of natural odors in ORN responses, we analyzed firing rates of 24 ORN types  in \emph{Drosophila } responding to a panel of 110 monomolecular odorants \cite{Hallem2006}.   We used this data to model responses to mixtures of odorants that are complex but sparse like natural odors. To do this, we constructed a firing rate ``response matrix'' $R$ whose entries specify the responses of each ORN to each monomolecular odorant.  We assumed that the ORN responses to odor mixtures are linear, which is a reasonable approximation at low concentrations \cite{mathis2016reading}.  This enabled us to define a complex mixture by a 110-dimensional composition vector $\vec{x}$ whose entries specify the concentrations (measured relative to \cite{Hallem2006}) of monomolecular odorants in the mixture.  The ORN firing rates $\vec{y}$ can then be modeled as linear combinations of responses to monomolecular odorants: $\vec{y} = R \, \vec{x}$.

To construct each mixture composition vector $\vec{x}$, we set a small number $K$ of its elements to be nonzero (where $K$ specifies the complexity of the mixture). The values of these nonzero entries were chosen randomly and uniformly between 0 and 2.  We then attempted to decode composition vectors ($\hat{x}$) from responses $\vec{y}$ using an efficient algorithm for decoding linearly-combined sparse composition vectors \cite{candes2009near,donoho2006compressed,candes2006robust}.  We deemed the result a failure if the average squared difference between components of the decoded ($\hat{x}$)  versus original ($\vec{x}$) composition vectors exceeded $0.01$,
 and defined \emph{decoding error}  as the failure probability over an ensemble of 500 odor mixtures $\{ \vec{x} \}$. This is a stringent criterion that we are using to quantify the accuracy with which mixture information is encoded in the ORN responses; there is no evidence to suggest that olfactory behavior requires this level of accuracy, nor do we assume that the brain uses this particular decoding scheme.   We checked that our findings are robust to different choices of failure threshold used to assess decoding error (Fig.~\ref{fig:suppfig2}).

 Fig.~2A shows the decoding error for varying mixture complexity $K$ and numbers of ORN types. Performance improves with increasing number of ORNs and decreasing mixture complexity. We compared the decoding error obtained from the measured ORN responses to two idealized alternatives: (1) a Gaussian random model, in which each ORN responds randomly to different odorants (with the overall mean and variance matched to data), and (2) a generalized ``labeled-line'' model, in which each ORN responds (with the same strength) to only five randomly-selected odorants.   The Gaussian random model would be an optimal strategy in the limit of many receptors and a large odor space \cite{candes2009near}, while the labeled line model is often considered to be a plausible interpretation of olfactory receptor responses.  The \emph{Drosophila} ORNs significantly outperform the labeled-line model and approach the performance of the Gaussian random model (Fig.~2C).  Quantitatively,  67\% of mixtures with 5 or fewer components drawn from 110 odorants can be accurately decoded from the responses of 24 receptors.   There are a staggering 100 million such mixtures.  Again, this is not to say that the fly brain attempts to reconstruct all of these odors with such an accuracy, but it does say that the receptors contain the necessary information.   Our theory also predicts that the olfactory representation of odors does not  depend on the details of how specific receptors respond to specific odors, but rather only depends on the broad distribution of responses across many receptors and many odors.  We tested this prediction by scrambling the  {\it Drosophila} response matrix (Fig.~2B) with respect to both odors and receptors and indeed found identical decoding performance (Fig.~2C).

 Our theory predicts that the olfactory code spreads information across all receptors, so that even weak responses are informative.   To test this comprehensively, we thresholded the {\it Drosophila} response matrix to keep only a fixed fraction of the strongest responses,  and then scrambled the odor identities for each receptor to create receptor responses with the same thresholded distribution. As predicted by our theory, as this fraction varied from $0$ to $1$,  decoding performance improved systematically (Fig.~\ref{fig:suppfig1}).

\subsection{The glomerular transformation increases disorder in response patterns}

Our theory suggests that disordered sensing --- in which a single receptor binds to many odorants, and a single odorant binds to many receptors --- is a powerful strategy for the olfactory system to employ.  However, {\it Drosophila} ORN responses  are noticeably structured and have a more clustered distribution of firing rates than, e.g., the Gaussian random model (Fig. 2B).     These correlations, perhaps arising from similarities between odorant binding sites or between receptor proteins, induce some order in receptor responses.  These responses are modified when receptors of each type converge to a second stage of processing in distinct glomeruli of the Antennal Lobe (analogously, the Olfactory Bulb in mammals).    There, a network of inhibitory interneurons reorganizes the receptor responses for transmission downstream  \cite{Friedrich2010Decorr}.  In the fly, the inhibitory network is well-described as effecting a divisive normalization \cite{olsen2010divisive, olsen2008lateral} that scales the responses of each ORN type in relation to the overall activity of all types (Appendix~B).   Applying this transformation to the {\it Drosophila} response matrix, we find that glomerular responses become more widely distributed and less correlated (Fig.~3A) than their ORN inputs.  This {\it densification} and {\it decorrelation} increases disorder.

 \begin{figure}
 \includegraphics[width=0.45\textwidth]{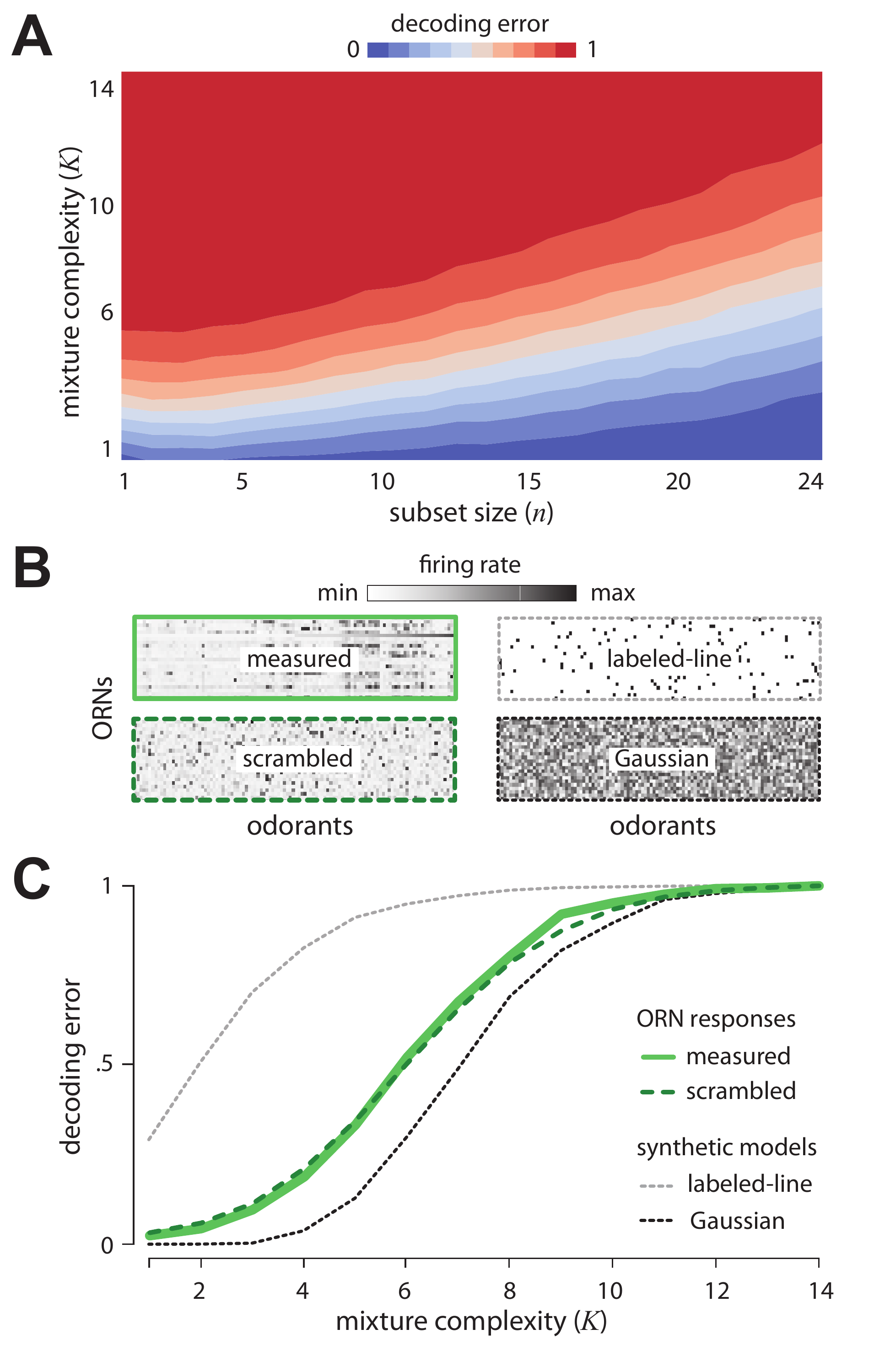}
 \caption{  {\bf Disordered sensing by ORNs enables accurate decoding of complex mixtures.}
 {\bf (A)}
 Error in decoding mixture composition from subsets of ORN responses, as a function of mixture complexity $K$ (i.e. number of mixture components) and ORN subset size $n$.
 Results are averaged over 500 odor mixtures of a given complexity, and 50 subsets of a given size.
 {\bf (B)}   Response matrices for {\it Drosophila} ORNs (measured and scrambled), labeled-line and Gaussian models (see text for details).
  {\bf (C)} Error in decoding complex mixtures from 24 ORNs as a function of mixture complexity $K$, shown for ORN responses (solid green), a scrambled version of ORN responses (dashed green), and two idealized models (the Gaussian random model, dashed black, and the labeled-line model, dashed gray).  Results are averaged over 500 odor mixtures of a given complexity. Results from scrambled, Gaussian, and labeled-line models are additionally averaged over 100 model instantiations.
 }
 \label{fig:Figure_2}
 \end{figure}

Does this increased disorder improve the representation of odor information? Because the divisive normalization  is nonlinear, we cannot, strictly speaking, use the aforementioned decoding algorithm to evaluate the information content of the glomerular representation.  However, we can instead create  an artificial benchmark in which mixtures $\vec{x}$ lead to responses $\vec{y}$ via
 $\vec{y}=R^{(2)}\vec{x}$, where $R^{(2)}$ represents a matrix of artificial glomerular responses obtained by transforming experimentally measured ORN responses to an odor panel in \cite{Hallem2006} via divisive normalization (see Appendix~B).
  Quantitatively, 67\% of mixtures with 7 or fewer components drawn from 110 odorants can be accurately decoded from the responses of 24 glomeruli, while similar accuracy was achieved for mixtures with only 5 components when decoding from ORNs (Fig.~3B).  Because the number of possible mixtures increases combinatorially with the number of mixture components, this is a substantial improvement.  A similar analysis shows that applying the divisive normalization
 to the labeled-line and Gaussian random models yields no improvement in decoding relative to the receptor stage  (Fig.~3B).

 As with decoding from ORNs, scrambling the responses over glomeruli and odors leads to identical decoding performance (Fig.~3B),  again suggesting that only the broad distribution of responses is important for the odor representation.  Weak responses remain informative; we again find that thresholding the response matrix degrades performance (Fig.~\ref{fig:suppfig2}).  Finally, we confirmed that our conclusions do not depend on details of the divisive normalization, but found, interestingly, that the experimentally-measured parameters \cite{olsen2010divisive} of this transformation minimize decoding error relative to other parameter choices (Fig.~\ref{fig:suppfig3}).

An alternative way of assessing the quality of a sensory representation is to ask how well it supports flexible associations between odors and valence.  To this end,
  we randomly labeled  mixtures ``appetitive'' or ``aversive'', and we trained a linear classifier  to identify these labels from ORN and fully nonlinear glomerular responses (Appendix~C).  Surprisingly, performance was poor (Fig.~3C), even though mixture compositions can be accurately decoded from these responses  (Fig.~2C \& 3B).
We conclude that although these first stages of processing retain nearly complete information about odor mixtures, this information is not readily usable for learning.

%----------------------------------------------------------- STAGE 3 ----------------------------------------------------------------
\subsection{Disordered projections reorganize odor information to facilitate flexible learning}

Although early stages of olfactory processing apparently do not support flexible learning, we know empirically that the representation at the third stage in the pathway {\it can} support such learning  (fly: \cite{mcguire2001role,heisenberg1985drosophila}; mammal: \cite{Choi2011}). How is odor information reorganized to achieve this?

In both insects and mammals, the transformation from the second to third stage of olfactory processing has two notable features: (i) expansive and disordered projections that distribute  odor information across a large number of cells \cite{caron2013random,sosulski2011distinct}, and (ii) nonlinearities that sparsify responses \cite{turner2008sparseMB,stettler2009representations}. As a result, an odor is represented by a sparse pattern of activity distributed broadly across cells in the third stage. We expect from general theory that this transformation should facilitate flexible associations between odor signals and valence \cite{cover1965geometrical,barak2013sparseness, babadi2014sparseness,luo2010generating}. Here, we propose that two additional sources of disorder -- densification achieved at earlier stages, and lack of structure in the connectivity patterns -- allow such associations to be learned from small groups of neurons drawn arbitrarily from within the population.

To test this, we simulated the responses of Kenyon cells in the Mushroom Body of the fly to odor mixtures (Fig. 4A).  We modeled each Kenyon cell as receiving inputs from $8$ glomeruli selected at random, reflecting empirical estimates \cite{caron2013random, litwin2017optimal} (interestingly, other choices yield worse performance; Fig.~\ref{fig:suppfig4}).  Connection weights were drawn uniformly between 0 and 1 (Fig. 4B, left). We  modeled long range inhibition by first removing the average response to an ensemble of odors, and then thresholding to eliminate weak responses (Appendix~D, \cite{luo2010generating}). This imposed a tunable level of sparsity in the population response. We fixed this sparsity to $15\%$ to match experimental estimates \cite{stettler2009representations, turner2008sparseMB}.  To assess learning, we generated responses to an ensemble of 5-component odor mixtures (as described above), and trained a linear classifier to separate responses into two arbitrarily-assigned classes  (Appendix~C). We defined {\it classification error} to be the fraction of mixtures that are incorrectly labeled by the classifier, averaged over 100 ensembles of mixtures and 100 labelings of each ensemble into appetitive/aversive classes.

We first compared classification from Kenyon cell responses (Fig. 4C) to that from responses of ORNs or glomeruli (Fig. 3C). To directly compare these different stages, we selected random subsets of $n = 160$ sparsely-active Kenyon cells. This ensured that any given odor would activate an average of 24 cells (0.15 $\times$ 160), matching the number of ORN and glomerulus types in our dataset.  We found that a linear classifier trained on Kenyon cell responses could categorize up to 300 mixtures with less than 10\% error (Fig. 4C), performing far better than a classifier trained on ORN or glomerular responses (Fig. 3C). In fact, even a much smaller population of $n = 80$ Kenyon cells (with an average of 12 active cells per odor) yielded better classification performance than the complete ORN or glomerular populations. Moreover, any arbitrary subset of a given size was equivalent (histogram inset of Fig. 4C). When we increased the number of cells used as a readout or decreased the average sparsity of responses, we found no improvement in classification (Fig.~\ref{fig:suppfig4}).

We then examined the role of disorder on classification performance. To do this, we separately removed each source of disorder (densification at the Antennal Lobe, and disordered projections from the Antennal Lobe to the Mushroom Body). To examine the role of the densification at the Antennal Lobe, we projected responses directly from the ORNs to the Mushroom Body, rather than passing responses through the transformation at the Antennal Lobe. To examine the role of disordered projection patterns, we introduced local structure in the projections from the Antennal Lobe to each subset of Kenyon cells in the Mushroom Body (Fig. 4B, right).  Within a given subset, we required that a fraction of Kenyon cells received preferential inputs from a fraction of glomeruli (in both cases, the fraction was taken to be 1/3). In doing so, we constrained the overall distribution of connection strengths to match those used to generate disordered connectivity (Appendix~E). This ensured that as a whole, each subset of Kenyon cells sampled all glomeruli, and any differences in performace were guaranteed to arise purely from differences in local connectivity patterns.

In the absence of neural variability, neither manipulation affected classification performance. However, both manipulations impacted performance in the presence of noise. To demonstrate this, we added proportional Gaussian noise of magnitude $\eta \sqrt{a r}$ to the firing rates $r$ of each ORN, where $\eta$ was drawn from a standard Gaussian and $a=.25$ controlled the coefficient of variation.  As expected, noise degraded performance (Fig. 4B,C).  Suprisingly, the impact of noise was worse when either of the two sources of disorder was removed, and even more when both sources were removed (Fig. 4C). Taken together, these results suggest that the disorder in the connectivity and the densification at the Antennal Lobe aids in learning flexible associations at the Mushroom Body.

\section{DISCUSSION}

We propose a new conceptual paradigm in sensory neuroscience: the use of {\it disorder} for building sensory representations that are accurate, compact, and flexible.  We argue that this paradigm explains the organization and function of the olfactory system, where disorder plays two key roles: ({\it i}) diffuse sensing by olfactory receptors serves to compress high-dimensional odor signals into compact neural representations, and ({\it ii}) densification followed by disordered expansion serves to reformat these representations for flexible learning. This paradigm exploits a key feature of natural odor signals---sparsity---to overcome a bottleneck in the limited number of olfactory receptor types. We used a combination of data and modeling to provide evidence for this paradigm in fly. Olfactory circuits in mammals show very similar anatomical and functional motifs, including broad receptor tuning \cite{saito2009odor} and apparently disordered projections to the cortex \cite{sosulski2011distinct}. This convergence between distant species suggests that disorder could provide a universal computational explanation for the architecture of early olfactory circuits.

\begin{figure}
\includegraphics[width=0.45\textwidth]{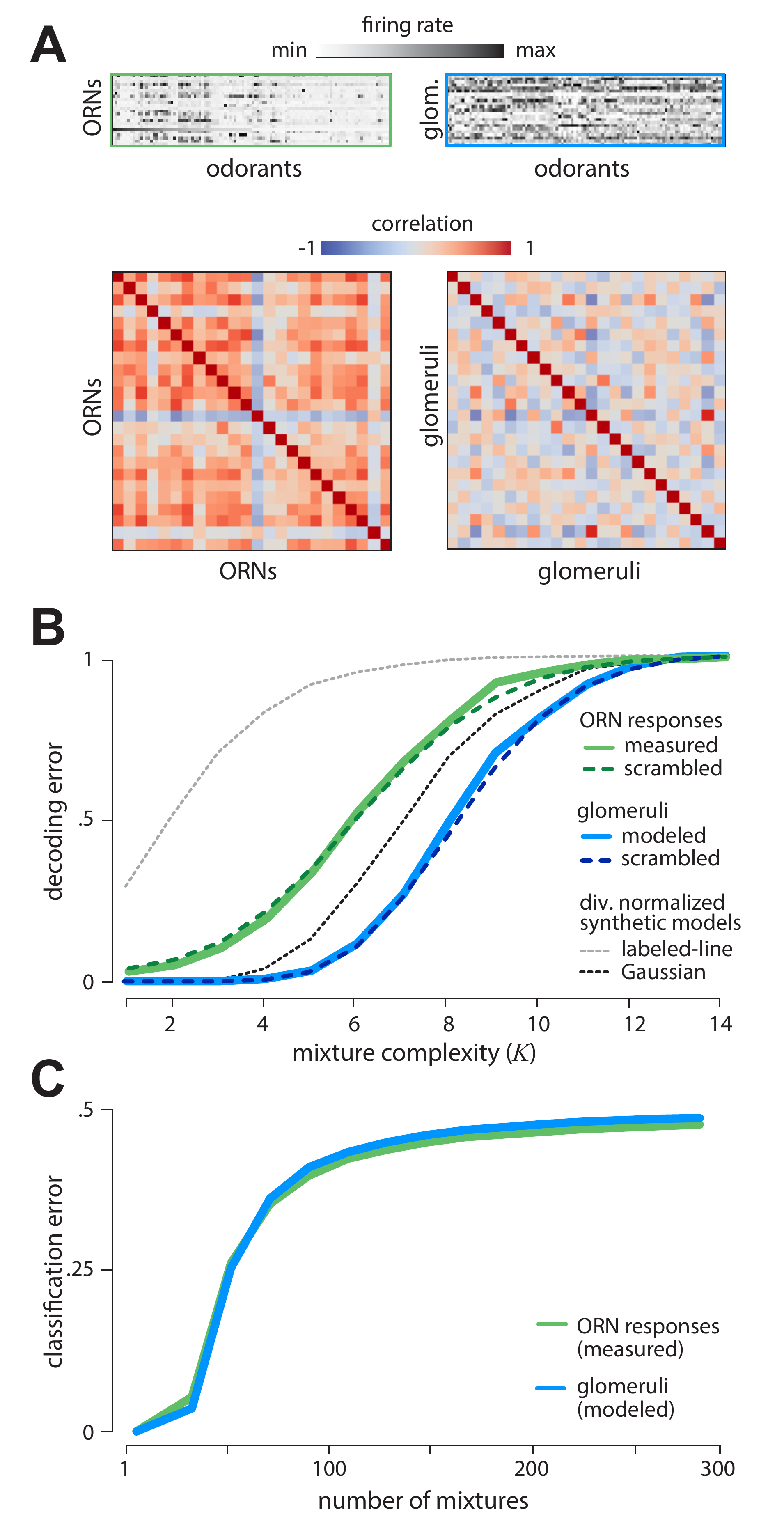}
\caption{  {\bf Divisive normalization in the Antennal Lobe increases disorder and decodability by {\it densifying} and {\it decorrelating} responses.}
{\bf (A)} Divisive normalization of ORN responses (top left) distributes responses more widely and densely over glomeruli (top right). Correlation coefficients between glomeruli (lower right) are much lower than between ORNs (lower left).
{\bf (B)} The error in decoding from glomeruli (solid blue) is much lower than from ORNs (solid green), and is unchanged by scrambling (dashed blue).  Divisive normalization has no effect on decoding error in the labeled-line model (dashed gray) or the Gaussian random model (dashed black). Results are averaged over 500 odor mixtures of a given complexity.
Results from scrambled, Gaussian, and labeled-line models are additionally averaged over 100 model instantiations.
{\bf (C)} Responses of ORNs and glomeruli are not readily usable for classification tasks. As shown, error in classifying mixtures from responses of ORNs (green) and glomeruli (blue) quickly approaches chance as the number of mixtures increases. Results are shown for two-class separability of 5-component mixtures, averaged over 100 different ensembles of odor mixtures, and 100 labelings into appetitive and aversive classes.
}\label{fig:Figure_3}
\end{figure}

{\bf The logic of olfactory receptors.}
Our theory predicts that general-purpose olfactory receptors should be selected for diffuse binding to many odorants, and not for the strong and specific binding often seen in biochemical signaling. An alternative view suggests that receptors should be adapted to bind selectively to molecules in particular odor environments or ecological niches \cite{carey2010odorant,zwicker2016receptor}.  These alternatives can be separated in experiments that measure the affinities of olfactory receptors to very large panels of odorants with varying ethological relevance. We predict that the typical receptor will have a diverse range of binding affinities across a broad array of odorants, with a statistically similar spread across molecules that both do and do {\it not} have immediate ethological importance.  Likewise, we predict that receptors in different species, even related ones, will typically have broadly different distributions of binding affinities, with similarities arising from biophysical constraints of olfactory receptors and not from properties of ecological niches. In addition, as a whole, the receptor repertoires of different species will show similar coverage across the space of odorants. This strategy resembles that of well-adapted immune repertoires, where different antibody distributions achieve similar coverage of the same pathogen landscape, as predicted theoretically \cite{mayer2015well} and observed in experiment \cite{davenport-2008,chain-2014,elhanati2014quantifying}.

\begin{figure}
\begin{center}
\includegraphics[width=0.45\textwidth]{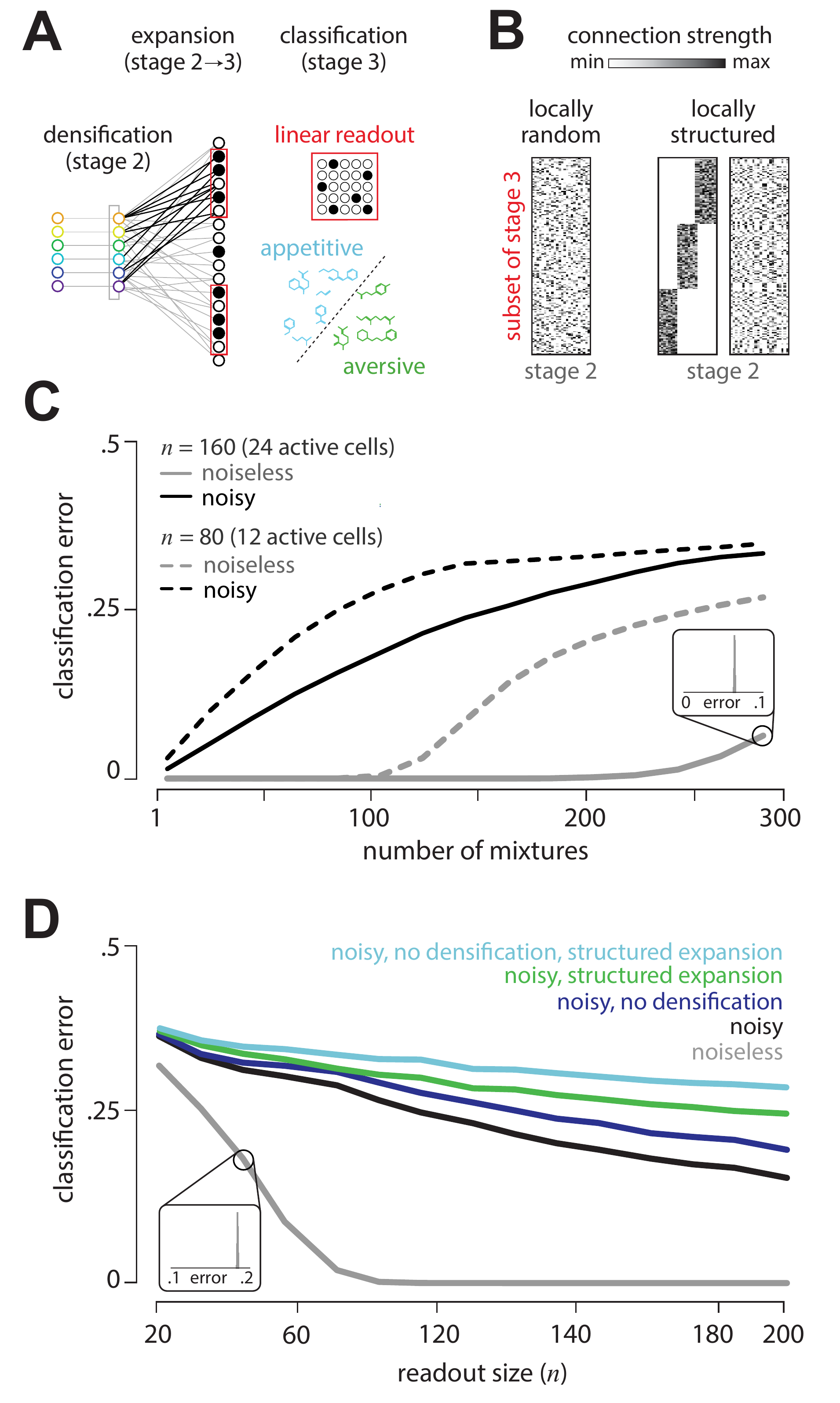}
\end{center}
\caption{  {\bf Disordered projections enable flexible learning in the presence of noise.}
{\bf (A)} Schematic. Small subsets of Kenyon cells exhibit sparse firing patterns in response to odor mixtures.  A linear readout neuron can learn to separate arbitrary classes of appetitive and aversive odor mixtures from these responses.
{\bf (B)} We generated random (left) and locally-structured (middle, right) projections from the Antennal Lobe (stage 2) to a subset of cells in the Mushroom Body (stage 3). Local structure was introduced by requiring that a fraction (1/3) of Kenyon cells receive inputs from a fraction (1/3) of all glomeruli (middle). When randomly permuted, the structure is no longer apparent (right).
{\bf (C)} A linear readout neuron can accurately classify mixtures from responses of small subsets of sparsely active Kenyon cells (here 15\% of cells respond).  Shown here is classification error for up to 300 different odor mixtures from subsets of 160 Kenyon cells (with 24 active cells; solid gray) or 80 Kenyon cells (with 12 active cells; dashed gray). All subsets of a given size produce nearly equivalent error (inset).  Noise (introduced in ORN responses; see text) degrades classification performance (black).
{\bf (D)} Classification error decreases as readout subset size increases (gray, noiseless; black, noisy). Removing disorder---either by removing densification (blue), introducing structure in the connectivity to the Mushroom Body (green), or both (cyan)---reduces robustness to noise.  All results are shown for two-class separability of 110 randomly-chosen 5-component mixtures. All results are averaged over 100 different ensembles of odor mixtures, 100 labelings into appetitive and aversive classes, and 100 instantiations of the projection pattern from the Antennal Lobe to the Mushroom Body.
}
\label{fig:Figure_4}
\end{figure}

{\bf The computational role of expansive and disordered projections.}
While this work provides evidence for the role of disordered sensing in the \emph{compression} of odor information, it also adds to a growing body of work on the computational role of \emph{expansion} via disordered neural projections. Expansive projections are known to make classification easier \cite{cover1965geometrical,barak2013sparseness,luo2010generating}, and the computational benefits of this expansion can be further improved by Hebbian learning \cite{babadi2014sparseness} and by sparse connectivity \cite{litwin2017optimal}.
 We have argued here  that the primary purpose of the expansion from the second to the third stage of olfactory processing is to reorganize a highly compressed representation of odors produced by disordered sensing by the receptors.   By contrast, other studies have proposed that this expansion could itself implement a form of odor signal compression \cite{krishnamurthyCOSYNE14, stevens2015fly}, or even a direct encoding of odor space \cite{zhang2016robust,kepple2016deconstructing} (in one case requiring unsupported assumptions about the mathematical relationship between the expansion and ORN responses \cite{zhang2016robust}). We found no evidence that expansive projections implement a form of compression, nor do we find evidence to support the direct representation of odor composition in Kenyon cell responses.
 Rather,
we found evidence that the expanded representation is organized to support
 flexible learning of categories \cite{gruntman2013integration,Choi2011}  from modest subsets of Kenyon cells.
 Anatomical evidence in fly indeed suggests that each olfactory readout neuron samples a only fraction of the  Mushroom Body \cite{schroll2006light} while still allowing formation of complex associations \cite{fiala2007olfaction}.
Our view is also consistent with abstract theory showing that sparsely firing binary neurons with ``mixed selectivity'' permit both discrimination between, and effective generalization from, complex overlapping binary inputs\cite{barak2013sparseness, rigotti2013importance}.    Our work can be viewed as additionally showing that {\it receptor} neurons with ``mixed selectivity'' effectively compress high dimensional sensory information, while subsequent ``mixed {\it sampling}'' of these responses reformats them for flexible learning by a simple readout.

{\bf Implications for behavior.}
Conceptually, our key idea is that disorder in the olfactory system is a fundamental adaptation to the intrinsic complexity of the world of smells.  We predict, distinctively, that odor information is distributed in both weak and strong responses across the entire ensemble of olfactory receptor types, and that this is important for complex discrimination tasks.  An alternative view suggests a ``primacy'' code where only the earliest or strongest responses are relevant for behavior \cite{kepple2016deconstructing}. We have shown (Fig. 2B and Fig.~\ref{fig:suppfig2}) that an encoding scheme that retains only the strongest responses contains much less information about complex mixtures than does a scheme that retains both strong and weak responses.  Because of this, we expect that our view can be separated from the primacy code in behavioral experiments that vary the complexity of discrimination tasks, e.g. by increasing the number of odors, the number of mixture components, and the degree of overlap between mixture components.  Given knowledge of responses to individual odorants, our theory quantitatively predicts the decline of behavioral performance with task complexity (e.g., Figs.~2,3,4). Likewise, our theory predicts how the relationship between behavioral performance and task complexity will vary as a function of information content in the olfactory pathway.  This information content can be experimentally manipulated by creating genetically-impoverished or enhanced receptor repertoires, optogenetically blocking inhibitory neurons in the Antennal Lobe to remove densification, or optogenetically activating Kenyon cells to simulate structured projection patterns from the Antennal Lobe.

{\bf Looking ahead.}
 Testing these predictions requires a movement away from simple paradigms involving small mixtures and pairwise discrimination, towards far more complex tasks that are reflective of life in the real world.  Methodologically, this shift has begun occurring in the study of vision.  We have argued here that in olfaction, this shift is even more critical --
the functional logic of the sense of smell can only be understand by taking into account the complexity of the real odor world.

\section{Appendices}

\subsection{Decoding odor composition}
  To  reconstruct $\vec{x}$ from measurements $\vec{y} = R\vec{x}$,
 we used the Iteratively Reweighted Least Squares (IRLS) algorithm \cite{chartrand2008iteratively} to find the  vector that minimizes the $L_1$ norm of $\vec{x}$ subject to the constraint
$\vec{y} = R\vec{x}$, with 500 maximum iterations and a convergence tolerance (in norm) of $10^{-6}$.

\subsection{Divisive normalization in the Antennal Lobe}
Lateral inhibition in the Antennal Lobe is believed to implement  a form of divisive normalization \cite{olsen2010divisive, olsen2008lateral,Friedrich2010Decorr}:
$R_{i}^{(2)}  =
R_{max}\cdot(R_{i}^{(1)})^{1.5}
/
\left[
\sigma^{1.5}+ (R_{i}^{(1)})^{1.5}+ (m\cdot{\displaystyle \sum_{i}}R_{i}^{(1)})^{1.5}
\right]$
where $R_{i}^{(1)}$ is the response of the $i$th ORN type, $R_{i}^{(2)}$ is the response of the $i$th glomerulus,  $\sigma$ parametrizes spontaneous activity, and $m$ controls the amount of normalization.  We use $R_{max} = 165.0$, $\sigma=10.5$, and $m=0.05$  \cite{olsen2010divisive}.   We constructed an artificial glomerular response matrix $R^{(2)}$ by applying this transformation separately to the ORNs responding to each of the 110 odorants studied in \cite{Hallem2006}.  Thus $R^{(2)}_{ij}$ represented the response of the $i$th glomerulus to the $j$th odorant.

\subsection{Linear classification}
To measure how well a particular odor representation (responses of ORNs, glomeruli, or Kenyon cells) facilitates learning flexible associations between odors and valences, we randomly split the representation of input mixtures into two classes and then trained a linear classifier (SVM with linear kernel \cite{pedregosa2011scikit}) to classify the inputs.

\subsection{Generating Mushroom Body responses}
 We took each Kenyon cell to have non-zero connection weights drawn uniformly between 0 and 1 with 8 randomly selected glomeruli (see Results). Then, following \cite{luo2010generating}, we took the input to the $i^{th}$ Kenyon cell, evoked by an odor with glomerular responses $\vec{y}$ in the Antennal Lobe, to be $h_{i} = \left< \vec{w_i}, (\vec{y} - \left<\vec{\mu}, \vec{y} \right>\vec{\mu}) \right>$, where $\left< \cdot, \cdot \right>$ is an inner-product, $\vec{w_i}$ is the vector of connection strengths, and $\vec{\mu}$ is the average Antennal Lobe response vector over all odors, normalized to unit length.   We chose a response threshold so that a fraction $f$ of neurons with inputs $h_i$ exceeding threshold are considered active, and normalized the thresholded responses so that the maximum firing rate is 5 Hz, on the order of the maximum observed Kenyon cell responses.  We averaged results over 100 random choices of connection strengths.   The global inhibition required in this model for generating the disordered responses observed in the Mushroom Body \cite{luo2010generating} could be implemented by the APL neuron which makes inhibitory connections to all the Kenyon cells

\subsection{Structured vs. random connectivity}
We constructed structured connectivity matrices between glomeruli in the Antennal Lobe and Kenyon cells in the Mushroom Body by reordering the columns of the corresponding random connectivity matrix so that the two matrices model synapses with the same connection strengths feeding into each Kenyon cell, but they sample different glomeruli. The reordering of the columns was done so that the structured connectivity matrix exhibited a block-diagonal structure as shown in Fig.~4B.   For analyses we chose the number of blocks to be 3.
 We then permuted the rows and columns of the structured connectivity matrix so that the underlying structure was not visible to the eye or to a casual analysis.

%%-----------------------------------------------------------------------------------------------------------------------------------------------
\subsection{Robust decoding from ORN and glomerular responses}

\begin{figure*}
\includegraphics[scale=0.52]{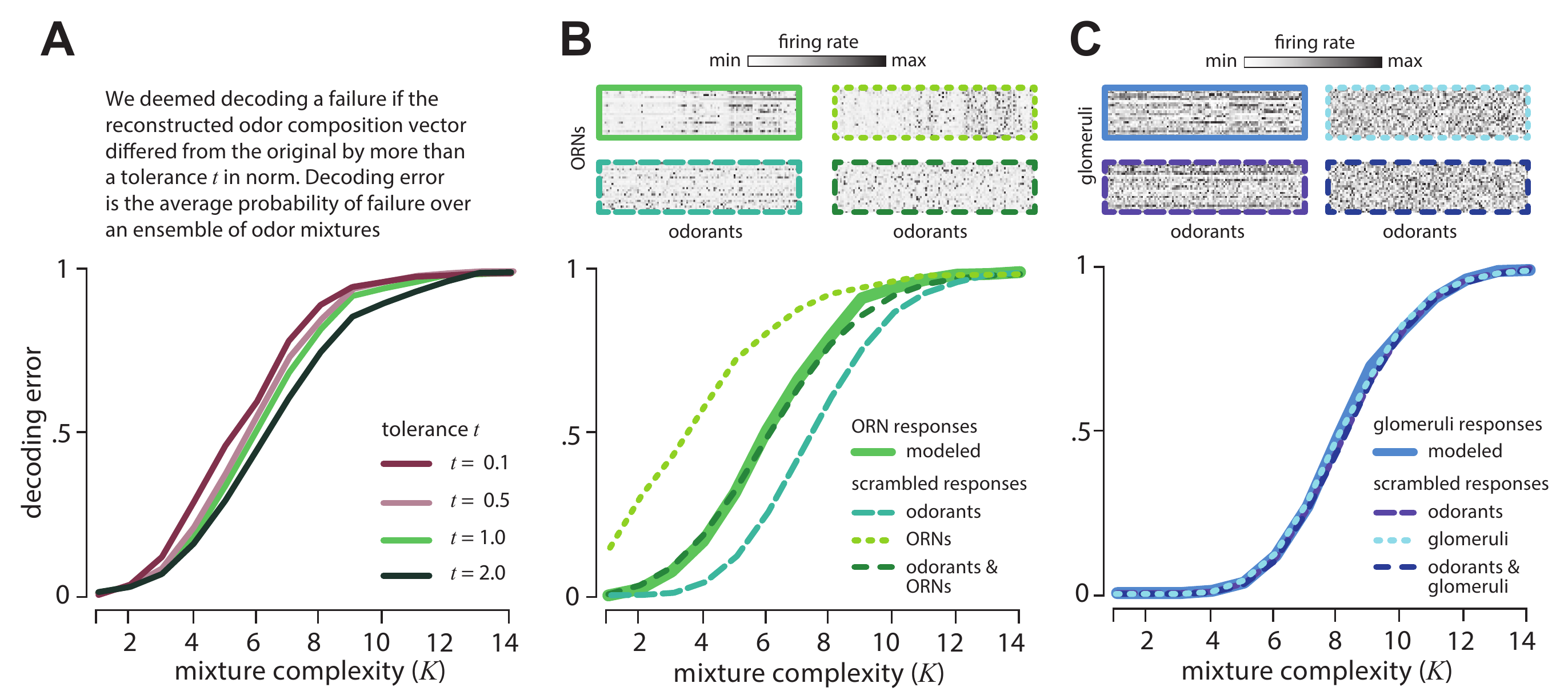}  
\caption{{\bf Odor decoding from {\it Drosophila} ORN responses is robust.}
(A) Decoding error is robust to ten-fold variations in odor reconstruction tolerance.  Mixture complexity = number of component odorants drawn from 110 possibilities.  (B) Decoding performance is unchanged after complete scrambling of the  {\it Drosophila} response matrix, because of opposite effects of scrambling receptors vs. odors.  Insets: Response matrices showing firing rates for 24 receptors (rows) responding to 110 monomolecular odorants (columns) without scrambling (solid green) and for models  randomly scrambling receptors, odorants, or both (dashed green).   (C) Decoding performance is unchanged after complete scrambling of the divisively normalized responses in the Antennal Lobe.  Separately scrambling receptors or odors also has no effect on performance.  Insets: Response matrices showing activity for 24 glomeruli (rows) responding to 110 monomolecular odorants (columns) without scrambling (solid blue) and after scrambling receptors, odorants, or both (dashed blue).   Results shown are averages over 100 iterations over model scrambled response matrices.  Decoding error is measured as the probability of decoding failure (see text) over an ensemble of 500 randomly chosen odor mixtures of a given complexity.
} \label{fig:suppfig2}
\end{figure*}

In the main text, we considered a simple linear model of the responses of 24 ORN types  in \emph{Drosophila } responding to  odor mixtures.  Specifically,  we extracted a firing rate matrix $R$ from the data in \cite{Hallem2006} (i.e. $R_{ij}$ is the response of receptor $i$ to odorant $j$), and we assumed that the response to a mixture could be written as a linear combination of responses to single odorants. We defined a mixture by the composition vector $x$ whose elements specify the concentration of individual odorants in the mixture.  The ORN firing rates $y$ could then be written as $\vec{y} = R \, \vec{x}$.  We then attempted to decode composition vectors $\vec{x}$ from responses $\vec{y}$ using the optimal algorithm of \cite{candes2009near,chartrand2008iteratively}.    We regarded the reconstruction as a failure if the average squared difference between components of the reconstructed odor vector and the original exceeded $0.01$.  Decoding error was defined as the failure probability over an odorant mixture ensemble.   This criterion for successful reconstruction is equivalent to saying that the reconstruction $ \hat{x} $  of the odor composition vector $\vec{x}$ fails if the norm of the difference $\| \vec{x} - \hat{x} \|$ exceeds a tolerance parameter  of $t = 1.1$ (here we used the fact that the odor composition vector $\vec{x}$ has 110 components).   To test the robustness of our conclusions we varied this tolerance parameter ten-fold, and found that the decoding error curves were largely unchanged  (Fig.~\ref{fig:suppfig2}A).
%(Supp. Fig.~\ref{fig:suppfig2}A).  
Qualitatively, we observed this robustness because the decoding of odors tends to either succeed very well, or fail very badly.   As a result, a broad range of criteria for defining a successful reconstruction will give similar measures of decoding error.

According to our general theory, and the results of \cite{candes2009near,candes2006near},  the quality of the olfactory code should not depend on the details of how  specific receptors respond to different odorants.  Rather, the key determinant should be the overall distribution of responses.   To test whether this is the case, we scrambled the receptor and odorant labels in the ORN response matrix  (top inset in 
%Supp. 
Fig.~\ref{fig:suppfig2}B), thus constructing an artificial response matrix with the same overall {\it distribution} of firing rates, but with no odor- or receptor-dependent correlations (second inset in 
%Supp. 
Fig.~\ref{fig:suppfig2}B).   We found that decoding performance was essentially identical when using the scrambled and unscrambled response matrices (
%Supp. 
Fig.~\ref{fig:suppfig2}B), consistent with the notion that the olfactory system seeks to employ disordered and unstructured sensing.   Interestingly, separate scrambling of the receptor labels and odor labels either improved or degraded the decoding, presumably because such scramblings removed correlations that were either detrimental or beneficial for decoding (
%Supp. 
Fig.~\ref{fig:suppfig2}B). These opposite effects compensated each other when the sensing matrix was fully scrambled.    We repeated this analysis after implementing a divisive normalization of ORN responses (see main text).  In this case, all scramblings left the decoding performance unchanged (
%Supp. 
Fig.~\ref{fig:suppfig2}C).  We thus conclude that after correlations are removed by divisive normalization, the overall distribution of responses is the sole determinant of the quality of the olfactory information representation.

\subsection{Weakly responding ORNs and glomeruli are informative}

\begin{figure*}
\includegraphics[scale=0.6]{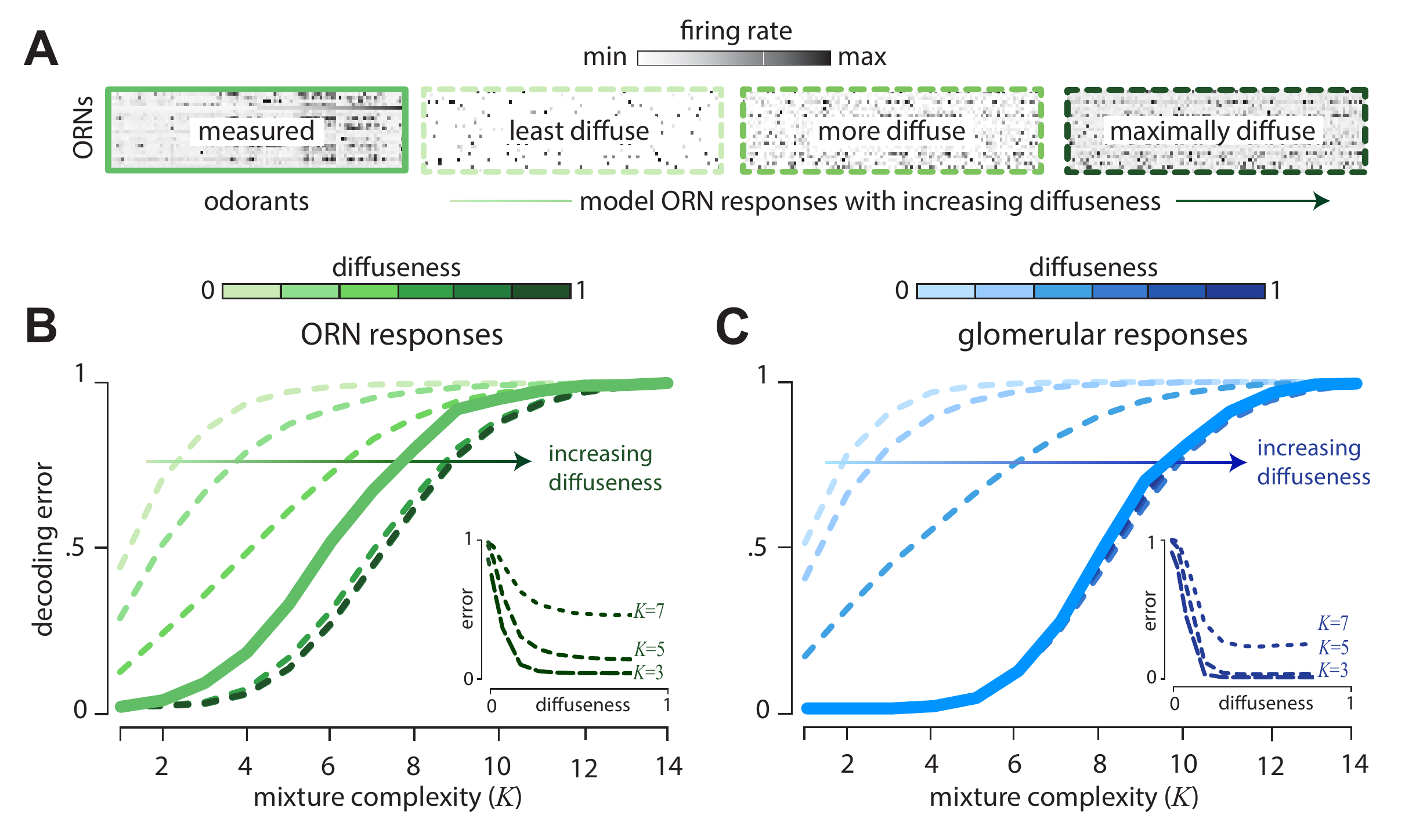}  \\
\caption{{\bf Weakly responding ORNs and glomeruli are informative about odor mixture composition}. (A) Firing rate response matrix measured from {\it Drosophila} ORNs (left, solid green), and for increasingly diffuse model response matrices (right, dashed green; ``diffuseness'' = fraction of largest responses kept).   Model responses are constructed by thresholding measured responses and then scrambling  the response matrix.  (B) Error in decoding from ORNs decreases systematically as diffuseness increases -- hence weak responses are informative.   Results shown as a function of mixture complexity ($K$ = number of odor mixture components). (C) ORN responses are divisively normalized to produce responses in the glomeruli of the Antennal Lobe (see Appendix~B).    Thresholding and scrambling these responses produces sensing models with different degrees of diffuseness.  Error in decoding from glomeruli decreases systematically as diffuseness increases.  Results shown are averages over 100 iterations over model response matrices for each degree of diffuseness.  Decoding error is measured as the probability of decoding failure over an ensemble of 500 randomly chosen odor mixtures of a given complexity.
} \label{fig:suppfig1}
\end{figure*}

Our theory predicts that the olfactory code is dispersed across all the receptors, so that even weak responses are informative.   To test this, we parametrized the fraction of largest responses that are deemed above threshold by a ``diffuseness parameter'' $f$.   We retained a fraction $f$ of the largest rank-ordered responses for each receptor, and we set the remaining values to zero.  A  diffuseness value of $f=1.0$ means we retain all responses, whereas a diffuseness value of $f=0.5$ means that we retained the strongest 50\% of all responses. We then created model response matrices for a given diffuseness value $f$ by randomly scrambling the thresholded receptor responses. 
%Supp.~
Fig.~\ref{fig:suppfig1}A shows the {\it Drosophila} ORN response matrix, along with model response matrices with increasing diffuseness.  
%Supp.~
Fig.~\ref{fig:suppfig1}B show  decoding error (definition in main text) as a function of mixture complexity $K$ ($K$ =  number of nonzero components in each mixture) for varying diffuseness. We see that decoding error decreases systematically as diffuseness increases, showing that weak receptor responses are informative about odor mixture identity.   The insets show decoding error as a function of the diffuseness parameter for fixed values of mixture complexity ($K=3,5,7$).  The results for the models with varying diffuseness are averaged over 100 randomly scrambled model response matrices.    
%Supp.~
Fig.~\ref{fig:suppfig1}C shows analogous results after applying divisive normalization to model responses in the glomeruli of the Antennal Lobe (see Appendix~B for details of this normalization).   The results show that weakly responding glomeruli are informative about mixture composition.

% -----------------------------------------------------------------------------------------------------------------------------------------------
\subsection{Optimal decoding from the Antennal Lobe}
\begin{figure*}
\includegraphics[scale=0.4]{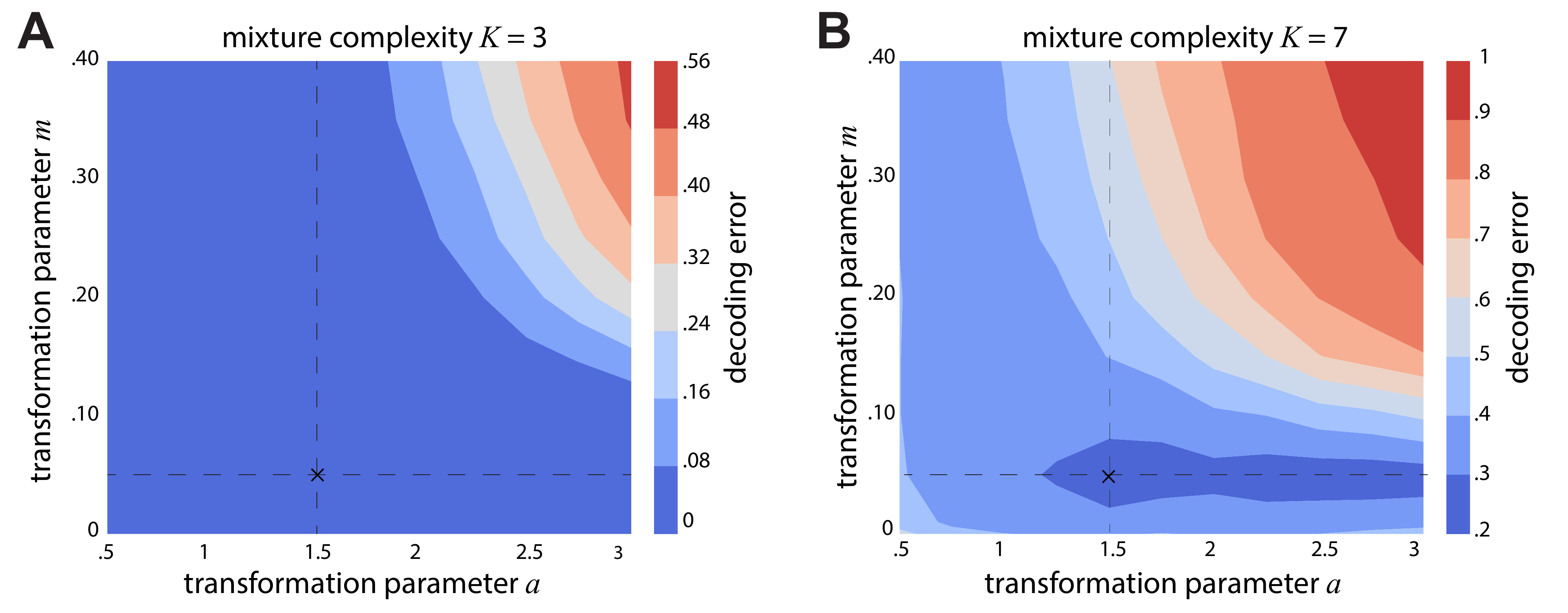}
\caption{ {\bf  The empirically determined divisive normalization in the Antennal Lobe is optimal for the measured ORN sensing matrix.}  Decoding error (see main text for definition) shown as a function of the exponent $a$, and the inhibition parameter $m$ in the divisive normalization carried out by the Antennal Lobe.  Left and right plots correspond to mixtures with $K=3$ and $K=7$ components drawn randomly from 110 odorants, respectively. The experimentally measured operating point is indicated by a cross in each plot ($m=0.05$ and $a=1.5$).   Decoding error (definition in main text) is averaged over 500 iterations of mixture ensembles of a given complexity.
 } \label{fig:suppfig3}
\end{figure*}

The inhibitory circuitry in the Antennal Lobe in {\it Drosophila} has been shown to perform a divisive normalization with the functional form  \cite{olsen2010divisive, olsen2008lateral}
\begin{equation}
R_{i}^{(2)}  =
\frac{
R_{max}\cdot\left(R_{i}^{(1)}\right)^{a}
}
{
\left[
\sigma^{a}+\left(R_{i}^{(1)}\right)^{a}+\left(m\cdot{\displaystyle \sum_{i}}R_{i}^{(1)}\right)^{a}\right] \, ,
}
\end{equation}
where $R_{i}^{(1)}$ is the response of the $i^{th}$ ORN type, $R_{i}^{(2)}$ is the response of the $i^{th}$ glomerulus,  $\sigma$ parametrizes spontaneous activity, and $m$ controls the amount of normalization.
 A fit to data in   \cite{olsen2010divisive, olsen2008lateral}  gave the parameters $R_{{\rm max}} = 165$, $\sigma=10.5$, $m=0.05$ and $a=1.5$.   In the main text,
we constructed an artificial glomerular response matrix $R^{(2)}$ by applying this transformation separately to the ORNs responding to each of the 110 odorants studied in \cite{Hallem2006}.  Thus $R^{(2)}_{ij}$ represented the response of the $i$th glomerulus to the $j$th odorant.
In the main text, we studied odor decoding in an artificial benchmark model in which mixtures $\vec{x}$ lead to responses $\vec{y}$ via  $\vec{y}=R^{(2)}\vec{x}$.
We tested how our results for decoding error (see definition in the main text and above) would be affected by changing the parameter $m$, which controls the amount of inhibition in the Antennal Lobe, or the exponent $a$, which controls the shape of the nonlinearity.
  In order to simplify our presentation, we study dependence on the parameters of the normalization for two values of mixture complexity: i) $K=3$, a value where odor reconstruction from Antennal Lobe responses with  experimentally-measured parameters is near perfect (see main text), and ii) $K=7$, a value where a similar reconstruction  starts to degrade.  (See main text for details regarding the construction of model odor mixtures of different complexities.)   We found that in both cases, the experimentally measured values of $m$ and $a$ led to the lowest decoding error (Fig. \ref{fig:suppfig3}).
  % (Supp. Fig. \ref{fig:suppfig3}).

%%-----------------------------------------------------------------------------------------------------------------------------------------------

\subsection{Mushroom Body classification error for mixtures}
\begin{figure*}
\includegraphics[scale=0.36]{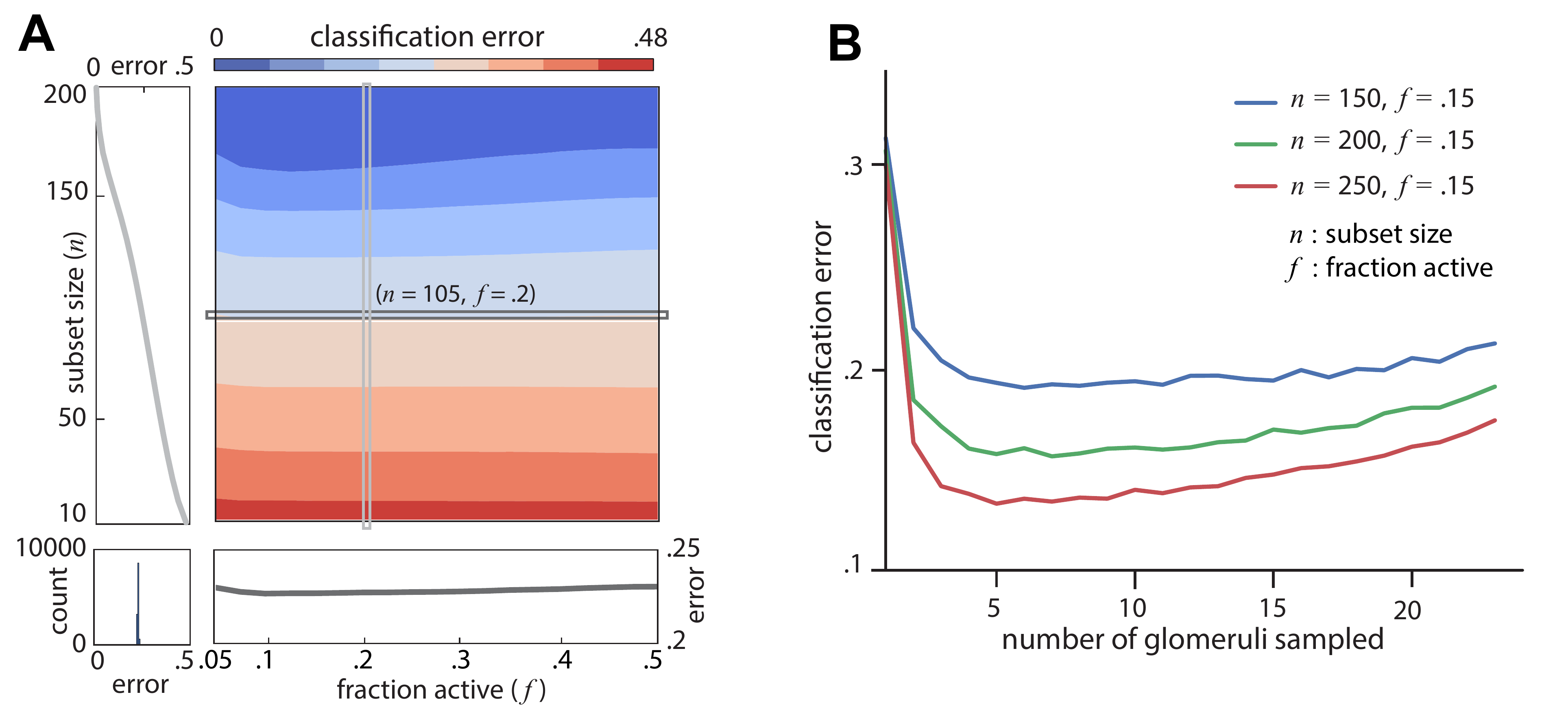}
\caption{A) Classification error from responses of model Kenyon cells in the Mushroom Body (MB) for arbitrarily separating 300 5-component mixtures into two classes as a function of  the readout size ($n$)
 and the fraction ($f$) of active neurons. The horizontal and vertical sections correspond to $n=105$ and
$f=0.2$, respectively (section shown in panels below and to the left, respectively).
Bottom left panel: histogram of classification errors for 10000 different
subsets of size $n=105$ and $f=0.2$. The narrowness of the histogram
shows that any two subsets of a given size are roughly equivalent for odor classification purposes.   B) Classification error at the Mushroom Body as a function of the number of glomeruli sampled by each Kenyon cell. Minimum error is found for sparse sampling of glomeruli.   All results shown are averages over 100 iterations over mixture ensembles, 100 labelings into appetitive/aversive classes, and 100 iterations over model connectivity matrices between the Antennal Lobe and Mushroom Body (each using a different instantiation of noise).  (See main text for details regarding the generation of connectivity matrices and noise.)
}
 \label{fig:suppfig4}
\end{figure*}

We studied the error in a 2-way classification task for 300 5-component mixtures with varying readout population sizes ($n$) and fraction of active Kenyon cells ($f$) in the Mushroom Body (details of classification procedure and task in the main text).
 For a given population size $n$, increasing the fraction of active neurons $f$ barely changes the classification performance (bottom panel of  
 %Supp. 
 Fig.~\ref{fig:suppfig4}A).
 The classification error with a given active fraction $f$ decreases with the number $n$ of neurons being read out  (left panel of  
 %Supp. 
 Fig.~\ref{fig:suppfig4}A). However, there is a law of diminishing returns -- excellent performance is achieved for relatively small $n$, and further increasing the population size makes little difference. The disordered projections from the Antennal Lobe to the Mushroom Body suggest that any subset of a given size should be statistically equivalent.   We tested this by comparing the classification error obtained from different  subsets of Kenyon cells.   The narrowness of the histogram of classification error for 10000 different populations ($n=105, f=0.2$) (lower left panel, 
 %Supp. 
 Fig.~\ref{fig:suppfig4}A) shows that any subset of a given size is indeed equally good at supporting flexible classification.

We also studied how the classification error depended on the number of glomeruli sampled by each Kenyon cell in the Mushroom Body. Figure \ref{fig:suppfig4}B shows the classification error as a function of the number of glomeruli sampled, for three different readout sizes. We see that the classification error initially decreases and then gradually rises as we increase the number of glomeruli sampled.  This indicates that there is an optimum for the number of sampled glomeruli. Recent work \cite{litwin2017optimal} has examined this question theoretically; here we show results with {\it Drosophila} data which are consistent with \cite{litwin2017optimal}.

\end{document}